\documentclass[sigplan,10pt,screen]{acmart}
\renewcommand\footnotetextcopyrightpermission[1]{}

\usepackage[hang,flushmargin]{footmisc}
\usepackage{booktabs} 
\usepackage{subcaption} 
\usepackage[normalem]{ulem}
\usepackage[export]{adjustbox}
\usepackage{enumitem} 
\usepackage{multicol}
\usepackage{setspace}
\usepackage[linesnumbered,boxed,lined,ruled,vlined,noresetcount]{algorithm2e}
\usepackage{mathtools} 
\usepackage{cleveref}
\usepackage{multirow}
\usepackage{array}
\usepackage{balance}
\usepackage{rotating}
\usepackage{outlines} 
\usepackage{bm} 
\usepackage{latexsym} 
\usepackage{xspace} 
\usepackage{color}
\usepackage{makecell} 
\usepackage{amsthm} 
\usepackage[zerostyle=d]{newtxtt} 
\usepackage[T1]{fontenc}
\usepackage{xtab,afterpage}
\usepackage{pifont} 
\usepackage{listings}
\usepackage{lstautogobble}
\usepackage{tabularx}
\usepackage{gnuplottex}
\usepackage{xcolor}
\usepackage{algorithmic}
\usepackage{graphicx}
\usepackage{breqn}
\usepackage{ragged2e}

\definecolor{codegreen}{rgb}{0,0.6,0}
\definecolor{codegray}{rgb}{0.5,0.5,0.5}
\definecolor{codepurple}{rgb}{0.58,0,0.82}
\definecolor{backcolour}{rgb}{0.95,0.95,0.92}
\definecolor{ballblue}{rgb}{0.13, 0.67, 0.8}
\definecolor{babyblue}{rgb}{0.54, 0.81, 0.94}
\definecolor{bleudefrance}{rgb}{0.19, 0.55, 0.91}
\definecolor{awesome}{rgb}{1.0, 0.13, 0.32}
\definecolor{brilliantrose}{rgb}{1.0, 0.33, 0.64}
\definecolor{brightpink}{rgb}{1.0, 0.0, 0.5}
\newtheorem{definition}{Definition}

\theoremstyle{nonumberplain}
\newtheorem{proof*}{Proof}
\theoremstyle{example}

\newcommand{\formal}[1]{${#1}$}
\newcommand{\sys}{\textsc{FM-Agent}\xspace}
\newcommand{\ccc}{CCC\xspace}
\newcommand{\vibetensor}{VibeTensor\xspace}
\newcommand{\vibeos}{VibeOS\xspace}
\newcommand{\vibedb}{Bespoke OLAP\xspace}
\newcommand{\sysspec}{\textsc{SysSpec}\xspace}

\AtBeginDocument{%
\hypersetup{
  colorlinks,
  linkcolor={purple!50!black},
  citecolor={red!50!black},
  urlcolor={blue!80!black}
}
}

\begin{document}

\title{\sys: Scaling Formal Methods to Large Systems via LLM-Based Hoare-Style Reasoning}

\renewcommand{\shortauthors}{}


\begin{abstract}
LLM-assisted software development has become increasingly prevalent, and can generate large-scale systems, such as compilers. It becomes crucial to strengthen the correctness of the generated code. However, automated reasoning for large-scale systems remains challenging due to code complexity. Hoare logic offers an approach to decomposing a large system into smaller components and reasoning about them separately (i.e., compositional reasoning). However, existing works still struggle to scale, because Hoare logic requires writing formal specifications for each function, imposing a heavy human burden. The problem is exacerbated when code is generated by LLMs, as developers lack a deep understanding of each function's expected behavior.

This paper presents \sys, the first framework that realizes automated compositional reasoning for large-scale systems. Leveraging LLMs, \sys introduces a top-down paradigm to automatically generate function-level specifications. Specifically, \sys derives the specification of a function from how its callers expect the function to behave, so the generated specifications can reflect the developer's intent of a function even if the implementation is buggy. Developers' intent is usually expressed in natural language, while existing verifiers only support formulas. Therefore, \sys generalizes Hoare-style inference to reason about functions against natural-language specifications. Finally, to confirm bug existence and explain bug causes, \sys automatically generates test cases to trigger potential bugs. In our evaluation, \sys successfully reasons about large-scale systems within 2 days, each of which has up to 143k LoC. These systems have already been tested by their developers, but \sys still finds 522 newly discovered bugs. These bugs can cause serious consequences, including system crashes and incorrect execution results.
\end{abstract}



\settopmatter{printfolios=true}

\author{Haoran Ding \qquad Zhaoguo Wang\textsuperscript{\ding{41}} \qquad Haibo Chen}
\affiliation{\institution{Institute of Parallel and Distributed Systems, Shanghai Jiao Tong University}\city{}\country{}}

\maketitle
{\renewcommand{\thefootnote}{\ding{41}}\footnotetext{Corresponding author.}}
\pagestyle{plain}

\section{Introduction}
\label{sec:intro}

LLM-assisted software development has become increasingly popular and can even generate systems with more than 100k LoC, such as the compiler \ccc~\cite{ccc}.
However, due to hallucinations of LLMs, the generated code may contain bugs.
Thus, reasoning about the correctness of such large-scale systems becomes crucial.
However, as codebases grow, existing automated reasoning techniques struggle to scale because system code combines complex control flow, rich state manipulation, and deep inter-procedural dependencies.

Compositional reasoning is a promising method to handle this problem.
The basic idea is to reason about each small component separately and compose the reasoning for each component to reason about the entire system.
Hoare logic~\cite{hoarelogic} realizes this idea based on the Hoare triple $\{P\}\ C\ \{Q\}$, which uses two formulas $P$ and $Q$ to define the formal specification of a code fragment $C$.
The pre-condition $P$ specifies what must hold before executing the code $C$, and the post-condition $Q$ specifies what will be ensured after $C$ terminates.
If the specification of each function is given, Hoare logic enables developers to reason about each function independently and compose the proof for each function to imply the correctness of the whole system.
This makes Hoare logic a strong foundation for compositional reasoning.

However, despite Hoare logic laying the foundation for compositional reasoning decades ago, existing techniques still struggle to fully realize this potential, even for sequential programs.
A critical bottleneck is the need for formal and human-written specifications.
It requires heavy human effort and deep understanding for large-scale systems.
For a long time, this bottleneck received little attention because writing proofs was seen as the heavier burden.
Recent work~\cite{chen2025atmosphere,zhang2025automan,sun2024anvil,zhou2024verismo,ironsync,armada,veribetrkv,komodo,ironfleet,verus2024sosp,dafny,Nelson2017hyperkernel,yv6,serval,Nelson2020bpf,klee,Helgi2018nickel,sosp2019vigor,yang2025autoverus,chen2024safe} has made major progress in automating proof generation.
As a result, specification generation is now one of the main challenges in automated reasoning for large-scale systems.
This challenge is intensifying in the era of LLM coding agents (e.g., Claude Code~\cite{claudecode}, GitHub Copilot~\cite{githubcopilot}, OpenAI Codex~\cite{openaicodex}, Cursor~\cite{cursor}).
Since systems are developed automatically rather than manually, developers often lack a deep understanding of the generated code.
Thus, it becomes more difficult to write accurate specifications.


This paper presents \sys, the first framework that enables automated compositional reasoning for large-scale systems.
It leverages LLMs to generate specifications at function granularity.
Each function can then be reasoned about concurrently under the principles of Hoare logic.
However, realizing \sys faces three key challenges:

\emph{{Challenge I: Specification generation must capture function behavior expected by the developer.}}
However, existing methods~\cite{chen2024safe} generate specifications from implementations, which may not capture the expected behavior,
because the implementation itself may be buggy or may fail to reflect original design intent.

\emph{{Challenge II: Developers' intent is naturally expressed in natural language, whereas existing verifiers reason only about formulas.}}
Specifically, developers usually write documents in natural language to describe the design of the whole system.
In the era of LLMs, prompts for code generation are also usually written in natural language.
However, existing verifiers support only formal specifications and cannot be applied.

\emph{{Challenge III: If verification via Hoare logic inference rules fails, verifiers cannot tell bug causes, which is important for bug fixing.}}
This challenge stems from the undecidability of program verification: no verifier can ensure both soundness and completeness.
Hoare logic ensures soundness without completeness and cannot even confirm the existence of bugs when verification fails.


To tackle these challenges, \sys proposes approaches based on three key insights:

\emph{{Insight I: LLMs can better capture the expected behavior of a function based on how its callers use it.}}
Then, \sys proposes a top-down paradigm to generate specifications.
To capture the design intent of each function, \sys uses LLMs to analyze the implementation of its caller to generate pre-conditions and post-conditions for the function.
For example, the arguments passed by the caller can help infer the pre-condition, and the subsequent code after invoking the callee can help infer its post-condition.
If a function is invoked by multiple callers, \sys merges the expected behavior inferred from each caller to generate a more comprehensive specification.
In this way, \sys will not be misled by buggy implementations.


\emph{{Insight II: LLMs are able to accurately predict the execution results of small code blocks.}}
For example, we generate random inputs and use LLMs to infer the outputs of functions from Microsoft VerusBench~\cite{verusbench}, a benchmark of Rust verification.
LLMs accurately predict the results for 98.6\% of the functions.
This insight enables \sys to directly perform reasoning against natural language specifications.
This is because LLMs understand both code semantics and natural language.
Specifically, \sys generalizes Hoare logic inference rules to support natural language pre-conditions and post-conditions.
Then, from the pre-condition in the specification, \sys uses LLMs to iteratively infer the post-condition of statements.
If the post-condition of the last statement in an execution path cannot imply the post-condition in the specification, \sys reports a potential bug.


\emph{{Insight III: LLMs can capture the correlation between the system entry input and internal function behavior.}}
Test cases that trigger bugs provide rich information for bug fixing.
Some commonly used methods, such as unit testing, focus on test cases for individual functions.
They cannot capture the correlation between system-entry inputs and bugs in internal functions.
This insight enables \sys to generate test cases at the system-entry level instead of the function level.
System-entry inputs are more intuitive for developers and less likely to obscure the root cause.
Thus, \sys iteratively generates test cases based on the reasoning process and runs test cases against the system to confirm whether the predicted bug is genuine.



In our evaluation, \sys reasons about multiple large-scale systems within 2 days, whose sizes range from 11k to 143k LoC.
The systems are automatically developed by various coding agents in different programming languages.
\sys still finds 522 newly discovered bugs even though their developers have already used various methods to test and fix these systems, such as unit testing, integration testing, differential checks, and multi-agent code review.
These bugs can lead to serious issues, such as system crashes and incorrect query results.

The source code of \sys is available at \url{https://github.com/fmagent-project/FM-Agent}.


\section{Hoare Logic Meets LLMs}
\label{sec:back}

This section introduces how Hoare logic enables compositional reasoning (\Cref{subsec:back:hoare}) and why existing works struggle to scale to large systems (\Cref{subsec:back:work}).

\subsection{Hoare-Style Compositional Reasoning}
\label{subsec:back:hoare}

The core of Hoare logic is the \emph{Hoare triple}:
\begin{equation*}
\{P\}\ C\ \{Q\}
\end{equation*}
It means that if $P$ holds before executing command $C$, then $Q$ holds after executing $C$.
For example, $\{x > 0\}\ y := x + 1\ \{y > 1\}$ is a valid Hoare triple.
At the function level, Hoare triples serve as \emph{contracts}: a function's pre-condition and post-condition together specify its expected behavior.
Callers must establish the pre-condition and may assume the post-condition, while the callee must ensure that its body satisfies the post-condition whenever the pre-condition holds.
This separation of concerns enables \emph{compositional reasoning}: each function can be verified independently against its contract, and the resulting local proofs compose into end-to-end guarantees for the entire program.
Thus, based on compositional reasoning, Hoare logic becomes a promising foundation for scalable verification.


Hoare logic provides many inference rules to realize compositional reasoning.
For example, the following rule means that if both Hoare triples above the line hold, then the triple below the line also holds.
\begin{equation*}
\frac{\{P\}\ F_1\ \{R\} \quad \{R\}\ F_2\ \{Q\}}{\{P\}\ F_1; F_2\ \{Q\}}
\end{equation*}
One then proves $\{P\}\ F_1\ \{R\}$ and $\{R\}\ F_2\ \{Q\}$ independently in parallel.
The rule then directly composes them to prove the larger code fragment $F_1; F_2$.
Although Hoare logic has provided a solid theoretical foundation, realizing compositional reasoning in large-scale systems remains difficult.
Hoare-style verification requires formal specifications for all functions, which requires deep domain expertise and manual effort.

\subsection{Existing Automated Reasoning Techniques}
\label{subsec:back:work}

We first introduce the limitations of traditional verification techniques and then discuss recent works that leverage LLMs to assist formal verification.

\paragraph{Automated and semi-automated verification.}
Symbolic execution~\cite{Nelson2017hyperkernel,yv6,serval,Nelson2020bpf,klee,Helgi2018nickel,sosp2019vigor} explores many feasible paths with symbolic inputs and checks desired properties on each path.
It can realize completely automated verification.
But it faces the issue of \emph{path explosion}, which means that the number of paths grows quickly with massive branches and loops.
It also cannot handle some program structures, such as \emph{unbounded loops}.
A loop is unbounded if the number of iterations cannot be determined statically, which is common in systems.
To handle the limitations of symbolic execution, semi-automated verifiers~\cite{chen2025atmosphere,zhang2025automan,sun2024anvil,zhou2024verismo,ironsync,armada,veribetrkv,komodo,ironfleet,dafny,verus,verus2024sosp} ask developers to provide annotations to assist the verification, which have stronger verification capabilities.
Developers need to manually write specifications for each function and some proofs, such as loop invariants and assertions.
Then, the verifier automatically generates and checks proof obligations based on SMT solvers.
However, manually writing these annotations requires substantial human effort, especially for large systems with many functions and complex logic.
This limits the scalability of these verifiers.

\paragraph{LLM-assisted verification.}
Recent work~\cite{chen2024safe,lahirie2024evaluating,yang2025autoverus} has explored using large language models (LLMs) to automatically reduce the manual effort of semi-automated verification.
For example, AutoVerus~\cite{yang2025autoverus} leverages LLMs to infer loop invariants from code.
This alleviates part of the annotation burden.
However, developers must still manually provide formal pre-conditions and post-conditions for each function, so these approaches cannot fully realize automated compositional reasoning.
Some other works~\cite{chen2024safe,lahirie2024evaluating} go further by using LLMs to generate function specifications as well.
However, they heavily rely on analyzing the program implementation, which reflects the implementation behavior rather than the expected behavior from developers.
The limitations of these works are further exacerbated in the era of LLM-based coding agents for two reasons.
First, LLM-generated code may contain subtle bugs that specification-from-implementation methods cannot detect.
Second, developers often lack a deep understanding of code they did not write, which makes it less feasible to manually write specifications.
In our scenario, we need to automatically generate specifications that reflect the design intent of developers regardless of implementation bugs.


\subsection{Insights and Our Approach}
\label{subsec:back:insight}

To handle the limitations of prior works, \sys is motivated by the following insights:

\paragraph{First, LLMs can better capture the expected behavior of a function based on how its callers use it.}
When writing caller functions, developers or LLMs usually have a clear intent of how callee functions should behave.
Thus, generating the specification of the callee based on the caller's implementation and specification can better capture the expected behavior.
For example, the invocation parameters from callers can help infer the callee's pre-condition.
The subsequent code manipulating the output of the callee in callers can help infer the callee's post-condition.
In this way, the generated specification is less likely to be misled by the buggy implementation of the callee itself.

\begin{figure}[t]
\begin{minipage}{1\linewidth}
\small
\begin{lstlisting}[escapeinside={(*@}{@*)},keywordstyle=\color{blue},morekeywords={pub,fn,let,match,None,Some,return}]
pub fn from_keyword(s: &str, gnu_extension: bool){
    let first = s.as_bytes()[0];
    // Fast reject by first character: keywords only start with these 16 chars.
    if !matches!(first, ... | b'l' | b'r' | ...) {
        (*@\label{fig:bg:example:line13}@*)return None;
    }
    match s {
        "auto"  => Some(TokenKind::Auto),
        ...
        _ => None,
    }
}
\end{lstlisting}
\normalsize
\end{minipage}
\begin{minipage}{1\linewidth}
\caption{An example of a buggy Rust function simplified from a C compiler~\cite{ccc} developed by Anthropic.}
\label{fig:bg:example}
\end{minipage}
\end{figure}

\begin{figure}[t]
\begin{subfigure}[t]{\linewidth}
\includegraphics[width=\linewidth]{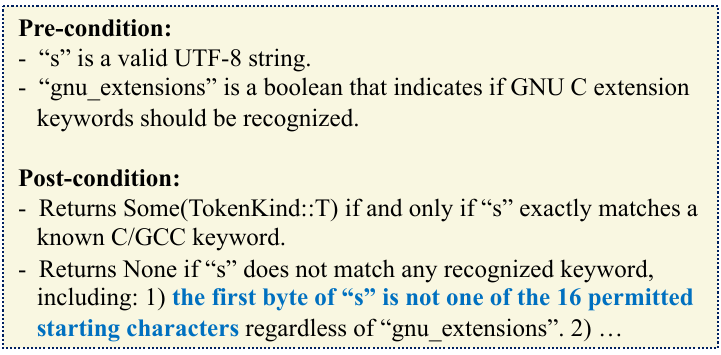}
\caption{The specification generated based on the implementation, which is buggy. The bolded part is incorrect.}
\label{fig:spec:buggy}
\end{subfigure}
\begin{subfigure}[t]{\linewidth}
\includegraphics[width=\linewidth]{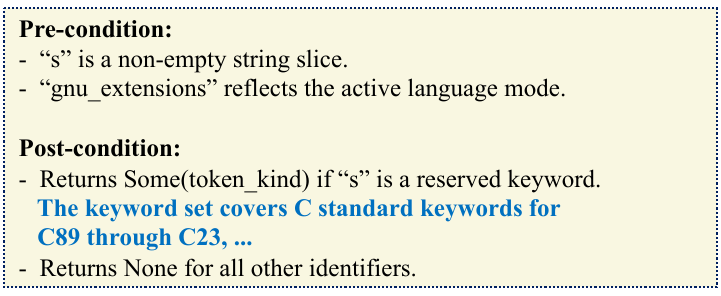}
\caption{The specification generated based on callers, which is correct. The bolded part fixes the error in the buggy specification.}
\label{fig:spec:correct}
\end{subfigure}
\caption{Specifications generated for the \texttt{from\_keyword} function in \Cref{fig:bg:example} using two methods.}
\label{fig:spec}
\end{figure}

For example, \Cref{fig:bg:example} shows a buggy function simplified from a C compiler called \ccc~\cite{ccc}, which is generated by Anthropic using Claude Code.
The function checks whether a string ``s'' is a keyword of C.
It fast rejects ``s'' by checking the first character of ``s'', and then checks whether the whole string matches any keyword in match arms.
However, it incorrectly rejects some keywords that start with characters other than the 16 characters in the fast rejection condition, such as ``nullptr''.
\Cref{fig:spec:buggy} shows the specification generated by Claude Code merely based on the implementation of \texttt{from\_keyword}.
It is clear that the specification is misled by the incorrect fast rejection code, which only considers keywords starting with the 16 characters as valid keywords.
As a result, the subsequent reasoning process based on this specification will miss the bug.
In contrast, \Cref{fig:spec:correct} shows the specification generated based on the caller of \texttt{from\_keyword}.
Compared with the specification in \Cref{fig:spec:buggy}, it does not claim that only keywords starting with the 16 characters are valid keywords.
Thus, it allows the subsequent reasoning process to find the bug.

\paragraph{Second, LLMs are able to accurately predict the execution results of small code blocks.}
For example, we perform a preliminary experiment based on all functions in VerusBench~\cite{verusbench}, a popular benchmark of Rust verification developed by Microsoft.
98\% of the functions in VerusBench have loops.
For each function, we randomly generate 3 inputs and use Claude Code to predict the execution results.
Claude Code correctly predicts the results for 98.6\% of cases (438 out of 444 cases), which demonstrates the capability of LLMs to accurately understand the semantics of small code fragments.
Thus, the insight enables stepwise Hoare-style reasoning in natural language: starting from the function's pre-condition, we use the LLM to infer the post-conditions of each small code block (e.g., a statement) and feed them as the next code block's pre-condition.
If the final post-condition of an execution path cannot satisfy the post-condition in the specification, we can report a potential bug.

\begin{figure}[t]
\begin{minipage}{1\linewidth}
\small
\begin{lstlisting}[keywordstyle=\color{blue},morekeywords={nullptr}]
int main() {
    int* ptr_a = nullptr;
    return 0;
}
\end{lstlisting}
\normalsize
\end{minipage}
\begin{minipage}{1\linewidth}
\caption{A test case generated by Claude Code to trigger the bug in \Cref{fig:bg:example}.}
\label{fig:bg:testcase}
\end{minipage}
\end{figure}

\paragraph{Third, LLMs can capture the correlation between the system entry input and internal function behavior.}
We perform a preliminary experiment based on 10 bugs in a C compiler called \ccc~\cite{ccc}.
We tell LLMs the reason why a function is buggy and ask LLMs to generate a C program to trigger the bug.
Even if the buggy function may be far from the compiler entry, LLMs still successfully understand the workflow and trigger all 10 bugs.
For example, the call chain from entry function to the buggy function \texttt{from\_keyword} in \Cref{fig:bg:example} contains 7 functions.
This insight is useful for bug fixing, because if we can directly generate system entry inputs rather than function-level inputs, developers can easily understand the root cause and fix the bugs.
Thus, \sys provides a function-level reasoning process for LLMs to generate system-entry inputs that trigger function bugs.

For example, the post-condition of line~\ref*{fig:bg:example:line13} in \Cref{fig:bg:example} is that the return value is None and ``s'' does not start with those 16 characters.
It cannot imply the post-condition in the specification shown in \Cref{fig:spec:correct}.
Based on the reasoning process, the bug validator successfully generates a C program shown in \Cref{fig:bg:testcase}.
The program uses a keyword ``nullptr'' that starts with a character other than the 16 characters, which cannot be recognized by \texttt{from\_keyword} due to the fast rejection code.

\section{\sys Overview}
\label{sec:overview}

\begin{figure*}[t]
\includegraphics[width=.98\textwidth]{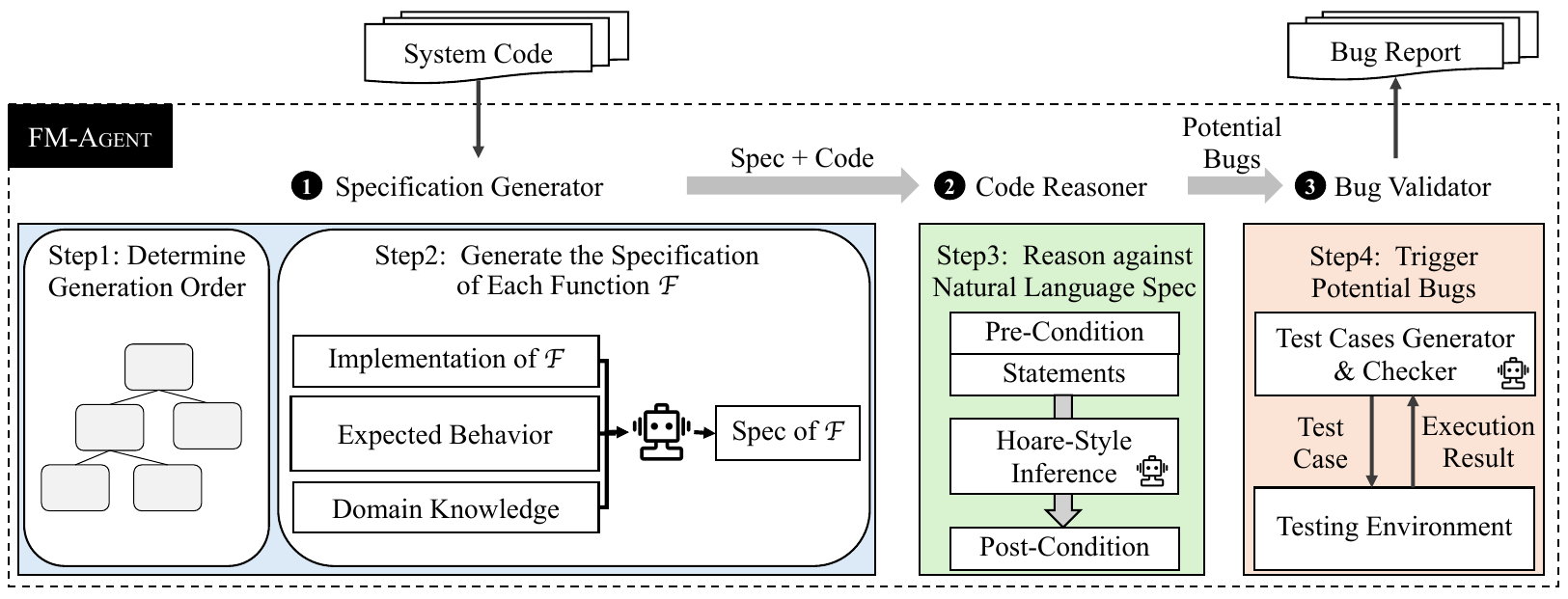}
\caption{The workflow of \sys.}
\label{fig:overview:workflow}
\end{figure*}

\Cref{fig:overview:workflow} shows the workflow of \sys, which automatically generates specifications and performs compositional reasoning.
The \emph{specification generator} (\Cref{sec:spec}) produces pre-conditions and post-conditions for each function.
It first determines the order in which specifications are generated.
This order is a partial order rather than a total order, which enables \sys to generate specifications for multiple functions concurrently.
As a result, the specification generator can scale to large-scale systems.
Then, for each function, the specification generator uses LLMs to derive the specification from three sources: the function's implementation, the expected behavior from callers, and domain knowledge (e.g., the C standard for compiler systems).

After generating specifications, the \emph{code reasoner} (\Cref{sec:reasoning}) checks whether each function's implementation is consistent with its specification.
It generalizes the inference rules of Hoare logic to operate over natural-language specifications and leverages LLMs to perform the reasoning.
Based on the principle of Hoare-style compositional reasoning, the code reasoner verifies each function concurrently and independently.
\sys currently supports only sequential functions, not concurrent functions.



Finally, the \emph{bug validator} (\Cref{sec:validate}) generates test cases and executes them to trigger potential bugs.
Note that it is impossible to completely avoid hallucinations of LLMs.
As a result, the code reasoner may produce false positives, i.e., cases where correct code is flagged as buggy.
Thus, we set a threshold to limit the number of attempts at test-case generation.
The bug validator reports it to developers only when the test cases successfully trigger the potential bug within this limit.

\section{Specification Generator}
\label{sec:spec}

Given a system codebase, the specification generator satisfies two key requirements.
First, for each function, the generated specification describes its expected behavior rather than the implementation steps.
Second, the specification generator fully exploits the concurrency potential of generating specifications for massive functions.
It allows \sys to be scaled to large codebases.

\subsection{Basic Idea}
\label{subsec:spec:idea}

Before introducing more details, we define expected behavior from caller functions as a new concept called \emph{expected specification}.
\begin{definition}[Expected Specification]
\label{def:expectedspec}
Assume a caller function $F_1$ invokes a callee function $F_2$.
The following formula represents that the expected specification of $F_2$ from $F_1$ is $\{P\} F_2 \{Q\}$.
\begin{equation*}
\begin{aligned}
F_1 \vdash \{P\} F_2 \{Q\}
\end{aligned}
\end{equation*}
Specifically, the formula means that: 1) the implementation of $F_1$ ensures that the program state immediately before the invocation of $F_2$ satisfies the condition $P$, and 2) the subsequent code of $F_1$ requires that the program state immediately after the invocation of $F_2$ satisfies the condition $Q$.
\end{definition}

\begin{figure}[t]
\includegraphics[width=\linewidth]{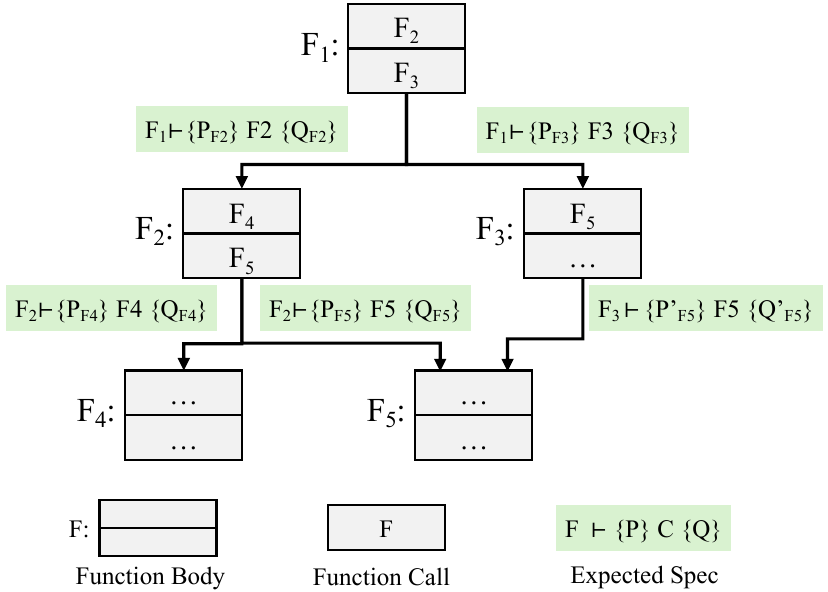}
\caption{An example of the top-down paradigm for specification generation. Each directed edge from function $F_i$ to function $F_j$ indicates that $F_i$ invokes $F_j$. The specification of $F_j$ is generated based on the expected specification from all its callers $F_i$, which is denoted as $F_i \vdash \{P\} F_j \{Q\}$ on the edge.}
\label{fig:spec:tree}
\end{figure}

The specification generator proposes a new top-down paradigm to automatically generate specifications.
\Cref{fig:spec:tree} shows the basic idea of the paradigm.
It starts from generating specifications for the entry functions (i.e., $\formal{F}_1$).
Using LLMs, \sys analyzes the function implementation and domain knowledge of the whole system to generate the specification for the entry function.
Furthermore, \sys also generates the expected specifications for all its callees $\formal{F}_2$ and $\formal{F}_3$, which are denoted by $F_1 \vdash \{P_{F2}\} F_2 \{Q_{F2}\}$ and $F_1 \vdash \{P_{F3}\} F_3 \{Q_{F3}\}$, respectively.
Then, \sys proceeds down the invocation graph.
Each function is processed after the specifications of all its callers are generated.
For each non-entry function, \sys generates its specification based on the expected specifications from its callers, the function implementation, and the domain knowledge.
Particularly, if a function is invoked by multiple caller functions, \sys combines the expected specifications from all its callers to generate its specification such that the generated specification can capture the expected behavior of the function in different invocation contexts.
For example, $\formal{F}_5$ has two callers $\formal{F}_2$ and $\formal{F}_3$.
Its specification is generated by combining $F_2 \vdash \{P_{F5}\} F_5 \{Q_{F5}\}$ and $F_3 \vdash \{P'_{F5}\} F_5 \{Q'_{F5}\}$.



\subsection{Deciding Specification Generation Order}
\label{subsec:spec:order}

We now describe how to determine the specification generation order, which is essential for enabling concurrent generation across large-scale systems.
Because some functions are independent of one another, the generation order is a \emph{partial} order rather than a total order: functions in the same layer of \Cref{fig:spec:tree} can be processed concurrently, as they depend only on the specifications of functions in earlier layers.
For instance, the specifications of $\formal{F}_2$ and $\formal{F}_3$ can be generated concurrently once the specification for $\formal{F}_1$ is ready.

\begin{algorithm}[t]
\setcounter{AlgoLine}{0}
\caption{Determining Specification Generation Order}
\label{fig:spec:order}
\footnotesize
    \let\oldgets\gets
    \renewcommand{\gets}{\textbf{:= }}
    \tt
    \textbf{Input:} A set of functions $\mathbb{F}$ in the codebase.\\
    \textbf{Output:} A sequence of function layers $[\mathbb{L}_1, \mathbb{L}_2, \ldots, \mathbb{L}_k]$, where each layer $\mathbb{L}_i$ is a function set.\\
    \underline{OrderPlanner}($\mathbb{F}$): \\
    \ \ $\langle V, E \rangle$ \gets ConstructCallGraph($\mathbb{F}$) \label{alg:order:callgraph} \\
    \ \ $SCCs$ \gets FindSCCs($\langle V, E \rangle$) \label{alg:order:scc} \\
    \ \ $\langle V’, E’ \rangle$ \gets CondenseGraph($\langle V, E \rangle$, $SCCs$) \label{alg:order:condense} \\
    \ \ $I$ \gets getIndegree($\langle V’, E’ \rangle$) \label{alg:order:topo} \\
    \ \ $\mathbb{Q}$ \gets $\{v \in V' \mid I(v) = 0\}$ \label{alg:order:initq} \\
    \ \ $k$ \gets $0$ \\
    \ \ \textbf{while} $\mathbb{Q} \neq \emptyset$ \textbf{do} \label{alg:order:while} \\
    \ \ \ \ $k$ \gets $k + 1$ \\
    \ \ \ \ $\mathbb{L}_k$ \gets $\mathbb{Q}$ \\
    \ \ \ \ $\mathbb{Q}$ \gets $\emptyset$ \\
    \ \ \ \ \textbf{foreach} $v \in \mathbb{L}_k$ \textbf{do} \\
    \ \ \ \ \ \ \textbf{foreach} $u$ $\in$ GetSucc($v, \langle V’, E’ \rangle$) \textbf{do} \\
    \ \ \ \ \ \ \ \ $I(u)$ \gets $I(u) - 1$ \\
    \ \ \ \ \ \ \ \ \textbf{if} $I(u) = 0$ \textbf{then} $\mathbb{Q}$ \gets $\mathbb{Q} \cup \{u\}$ \\
    \ \ \textbf{return} $[\mathbb{L}_1, \mathbb{L}_2, \ldots, \mathbb{L}_k]$ \\
    \let\gets\oldgets
\normalsize
\end{algorithm}

\Cref{fig:spec:order} presents the order-planning algorithm.
Given a set of all functions $\mathbb{F}$ in the system, the algorithm produces a layered sequence $[\mathbb{L}_1, \mathbb{L}_2, \ldots, \mathbb{L}_k]$, where each $\mathbb{L}_i$ is a set of functions whose specifications can be generated concurrently.
The sequence guarantees that every caller of a function in $\mathbb{L}_i$ belongs to some layer $\mathbb{L}_j$ with $j \leq i$; equality holds only when the caller and callee invoke each other.
Therefore, \sys can follow the top-down paradigm by processing the layers in order, from $\mathbb{L}_1$ to $\mathbb{L}_k$.
The algorithm proceeds in four steps.
First, \texttt{ConstructCallGraph} constructs a function call graph $\langle V, E \rangle$ from $\mathbb{F}$ (\Cref{alg:order:callgraph}), where each node represents a function and a directed edge from $F_1$ to $F_2$ indicates that $F_1$ invokes $F_2$.

Second, \texttt{FindSCCs} identifies all strongly connected components (SCCs) in the call graph (\Cref{alg:order:scc}).
An SCC is a maximal subgraph in which every node is reachable from every other node.
This step is necessary because recursive function invocation can construct cycles.
The specification of functions in the same SCC will be generated simultaneously, as their specifications are interdependent.

Third, \texttt{CondenseGraph} condenses the call graph by collapsing each SCC into a single node, yielding a directed acyclic graph $\langle V', E' \rangle$ (\Cref{alg:order:condense}).
Because functions in the same SCC will be specified simultaneously, the algorithm treats them as one unit for ordering purposes.
The algorithm replaces all cross-SCC edges with edges between the corresponding condensed nodes.

Finally, it performs a layered topological sort on the condensed DAG (\Cref{alg:order:topo}--\Cref{alg:order:while}).
It first calls \texttt{getIndegree} to compute the in-degree of every node in $\langle V', E' \rangle$ (\Cref{alg:order:topo}) and initializes a set $\mathbb{Q}$ with all zero-in-degree nodes (\Cref{alg:order:initq}).
The initial $\mathbb{Q}$ represents all entry functions without callers in the system.
In each iteration, the current $\mathbb{Q}$ becomes the next layer $\mathbb{L}_k$.
The algorithm then calls \texttt{GetSucc} to enumerate each successor of nodes in $\mathbb{L}_k$ (i.e., their callees), and decrements their in-degrees.
If any successor's in-degree drops to zero, it is added to the next $\mathbb{Q}$ for processing in the next iteration.
The process repeats until $\mathbb{Q}$ is empty, yielding $[\mathbb{L}_1, \mathbb{L}_2, \ldots, \mathbb{L}_k]$.

During specification generation, \sys processes the layers from $\mathbb{L}_1$ (the entry functions) to $\mathbb{L}_k$.
This ensures that, when generating the specification for a function, the specifications of all its callers outside its SCC (if any) have already been generated.
Functions within the same layer are generated concurrently.
For functions that share an SCC, they are in the same layer.
\sys generates their specifications simultaneously.
\Cref{subsec:spec:topdown} discusses how specifications are generated for functions within the same SCC.

To further improve efficiency, \sys leverages batching and phase-level concurrency.
The specification generator can batch the specification generation for functions in the same layer.
Assume the batch size is $N$, the specification generator groups the prompts for up to $N$ functions in the same layer into a single batch and sends them to LLMs together.
This can reduce network round trips and improve overall throughput.
The batch size $N$ can be tuned for different codebases and models.
Too large a batch may increase latency and degrade inference quality, while too small a batch may underutilize available concurrency.
Phase-level concurrency means that some systems naturally consist of multiple self-contained phases, where each phase can be processed independently and concurrently.
This is because, for each phase, most functions only invoke functions in the same phase.
For example, a compiler can be split into multiple self-contained phases, such as lexing, parsing, IR generation, optimization, and executable file generation.

\subsection{Generating Caller-Driven Specifications}
\label{subsec:spec:topdown}


For each function, the specification is derived by LLMs from three sources: 1) domain knowledge of the system, 2) expected specifications from callers (\Cref{def:expectedspec}), and 3) the function implementation.
For entry functions, \sys uses only sources (1) and (3); for internal functions, \sys uses all three sources.

Domain knowledge helps LLMs understand the expected behavior of the system, so they can generate accurate specifications, especially for new kinds of systems.
For example, when reasoning about a C compiler, the domain knowledge about the C standard can help LLMs understand what C programs are valid.
If the system is common, domain knowledge can be summarized from standards.
If the system is new, developers can write documents that provide the necessary domain knowledge for LLMs.
However, such documents may be very long, so providing all of them to LLMs is often impractical because of context window limits.
We also observe that systems are usually modular, and different modules require different domain knowledge.
For example, in a C compiler, the lexer and IR modules rely on different knowledge.
Therefore, before specification generation, we use LLMs to analyze system components and split domain knowledge into separate files by component relevance.
When generating the specification for functions in a component, \sys only provides the domain knowledge that is relevant to the component to LLMs.

For each function, after generating its own specification, \sys also generates expected specifications for its callees.
These expected specifications, defined in \Cref{def:expectedspec}, are natural-language pre/post conditions that guide the next layer.
If a function is invoked by multiple callers, \sys combines all caller-provided expected specifications to construct a comprehensive specification.
The basic idea of combination is shown as follows.
\begin{equation*}
\begin{aligned}
\frac{F_1 \vdash \{P_1\} F \{Q_1\} \quad \ldots \quad F_n \vdash \{P_n\} F \{Q_n\}}{ P_F \coloneq (P_1 \lor \ldots \lor P_n) \quad Q_F \coloneq (Q_1 \land \ldots \land Q_n)}
\end{aligned}
\end{equation*}
Specifically, assume a function $F$ is invoked by $n$ callers $F_1, \ldots, F_n$.
Given the expected specifications from all callers, we can combine them to derive the specification for $F$.
The pre-condition $P_F$ is the disjunction $P_1 \lor \ldots \lor P_n$, because $F$ may be invoked from the context of any caller $F_i$.
The post-condition $Q_F$ is the conjunction $Q_1 \land \ldots \land Q_n$, because $F$ should satisfy the expected post-conditions from all of its callers simultaneously.
Note that the formula just illustrates the basic idea of combination.
The actual combination is implemented by LLMs rather than directly concatenating sub-formulas by $\lor$ and $\land$, because $P_i$ and $Q_i$ are expressed in natural language rather than formulas.
Combining natural-language specifications is more complex than combining formal specifications.
It may require understanding the semantics of the specifications and rephrasing them if there are redundancies or potential conflicts.

Although a function implementation may be buggy, it can still provide useful information.
In particular, entry functions have no callers, so their specifications are mainly based on domain knowledge and implementation.
The implementation is also useful for functions in the same call-graph cycle, which invoke each other.
When generating specifications for a function in a cycle, \sys lacks the expected specifications from its callers in the same cycle.
This is because these functions are placed in the same layer, as described in \Cref{subsec:spec:order}.
Thus, for a function in a cycle, the implementations of its callers in that cycle help LLMs infer the function's expected behavior from their perspective.

Based on the above method, the specification generator finally generates a separate file for each function, which contains three parts.
The first part is the function specification, including the pre-condition and post-condition in natural language.
The second part is the expected specifications for its callees, which are also in natural language.
The third part is the function body.
\sys will reason about each function against the specification concurrently (\Cref{sec:reasoning}).

\subsection{Top-Down vs. Bottom-Up}
\label{subsec:spec:discuss}

Traditionally, when manually writing specifications, some developers prefer a top-down manner while others prefer a bottom-up manner.
In the bottom-up manner, developers first write specifications for callee functions and then for their callers.
There is no clear consensus on which manner is better for manually written specifications.
In this subsection, we first explain why some developers favor the bottom-up manner when \emph{manually writing specifications}, and then discuss why the top-down manner is more suitable when \emph{automatically generating specifications}, especially for LLM-generated code.

Developers who manually write specifications may prefer the bottom-up manner for two reasons.
First, because systems are traditionally developed by hand, developers usually have a clear understanding of the semantics of each function and the system as a whole.
With this understanding, they can capture the intended function behavior regardless of whether they work top-down or bottom-up, even when the implementation contains bugs.
Second, the bottom-up manner avoids the need to directly write a formal specification for the entire system at the outset, which can be difficult because formal specifications are precise and may involve many details.
Instead, the specification of a caller function can reuse the formulas or data structures already defined in the specifications of its callees, reducing the overall burden.
In particular, recent layer-based verification methods~\cite{certikos} make bottom-up specification writing especially natural.
These methods split the system into multiple layers, where upper-layer functions invoke lower-layer functions, and verify each layer in turn.
Because sub-formulas and data structures from a lower-layer specification are typically reused when reasoning about the upper layer, writing specifications in a bottom-up manner aligns well with this verification strategy.

However, when we aim to automatically generate specifications or the code is produced by LLMs, the two reasons above no longer hold, and the top-down manner becomes more suitable.
First, when the code is written by LLMs rather than humans, developers lack a deep understanding of the expected behavior of each function, so the first reason no longer applies.
Second, the implementation may contain bugs, which can mislead specification generation under the bottom-up manner: because callee specifications are generated before those of their callers, they are derived primarily from potentially buggy implementations.
In contrast, the top-down manner generates caller specifications first, capturing expected behavior from the perspective of the whole system even when implementations are buggy.
Third, automatically generated specifications are typically expressed in natural language, which tolerates fewer details than formal specifications; thus the second reason no longer applies either.
With the top-down manner, specifications can be generated starting from the entry functions and progressively introducing more details for inner functions as needed.

\section{Code Reasoner}
\label{sec:reasoning}

For each function, the input of the code reasoner includes the function implementation, its natural language specification, and the expected specifications for all of its callees.
The code reasoner reasons about whether the implementation of the function is consistent with its specification via Hoare-style inference.
During the reasoning process, we assume that expected specifications of its callees are satisfied, which allows us to focus on the current function and realize compositional reasoning.

\subsection{Basic Idea}
\label{subsec:reason:natural}

As described in \Cref{subsec:back:insight}, \sys leverages LLMs to reason about the implementation directly against natural language specifications.
The basic idea is to generalize Hoare-style inference rules to generate natural language post-conditions for statements.
Then, the code reasoner checks whether the post-condition of the last executed statement in each execution path implies the post-condition of the function, such as a return statement.
If the check fails, it indicates that the implementation of the function may be inconsistent with its specification, which will be reported as a potential bug.
The report includes the specific statement where the violation occurs and the reasoning process.

\begin{figure}[t]
\begin{minipage}{1\linewidth}
\small
\begin{lstlisting}[escapeinside={(*@}{@*)},keywordstyle=\color{blue},morekeywords={pub,fn,let,match,None,Some,return}]
// {"s" is a non-empty string slice. "gnu_extensions" reflects the active language mode.}
let first = s.as_bytes()[0];
// {"s" is a non-empty string slice. "gnu_extensions" reflects the active language mode. "first" is the first byte of s.}
if !matches!(first, ... | b'l' | b'r' | ...) {
    // {"s" is a non-empty string slice. "gnu_extensions" reflects the active language mode. "first" is the first byte of s. "first" is not in ('l', 'r', ...).}
    return None;
    // {"s" is a non-empty string slice. "gnu_extensions" reflects the active language mode. "first" is the first byte of s. "first" is not in ('l', 'r', ...). The function return value is None.}
}
\end{lstlisting}
\normalsize
\end{minipage}
\begin{minipage}{1\linewidth}
\caption{The reasoning process of the function \texttt{from\_keyword} in \Cref{fig:bg:example}. For brevity, we only show the code fragment that contains the bug. Every comment is the post-condition of the preceding code and the pre-condition of the following code.}
\label{fig:reasoner:example}
\end{minipage}
\end{figure}

For example, \Cref{fig:reasoner:example} shows the reasoning process of the function \texttt{from\_keyword} in \Cref{fig:bg:example}.
The code reasoner generates the post-condition for each statement based on the pre-condition and the statement itself.
The pre-condition of the first statement comes from the specification (\Cref{fig:spec:correct}).
The post-condition of the first statement adds a new proposition: \texttt{first} is the first byte of \texttt{s}.
Then, the if statement further constrains \texttt{first} to be outside a character set.
\sys then checks whether the post-condition of the return statement entails the post-condition in the specification, which is ``if \texttt{s} is a keyword, return \texttt{Some}; otherwise return \texttt{None}''.
The entailment fails because the inferred condition is insufficient to establish the ``otherwise return \texttt{None}'' branch.
Accordingly, \sys reports the return statement as a potential violation site.

\subsection{Post-Condition Inference}
\label{subsec:reason:inference}

Now we introduce the details of inferring the post-condition, which is derived from the pre-condition and the statement itself.
The pre-condition of the first statement is the pre-condition in the function specification.
The pre-condition of the subsequent statements is derived from the post-conditions of the previous statements.
With the pre-condition, \sys generalizes different Hoare logic inference rules to infer the post-condition for different types of statements.
For simplicity, we only explain the inference method for some important types of statements.
The inference method of other statements is similar to Hoare logic inference rules, which are omitted.


\paragraph{Branch statement.}
The post-condition is the disjunction of the natural language post-conditions of all branches.
This is consistent with the classic Hoare-style inference rules for if-else statements, except that the pre-conditions and post-conditions are in natural language rather than formulas.
\begin{equation*}
\begin{aligned}
\frac{\{P \land C\} C_1 \{Q_1\}\quad \{P \land \neg C\} C_2 \{Q_2\}}{\{P\} if (C)\,\, C_1\,\, else\,\, C_2 \{Q_1 \lor Q_2\}}
\end{aligned}
\end{equation*}

\paragraph{Loop statement.}
For a loop statement, \sys first generates the post-condition of the loop via LLMs.
Then, it tries to generate a loop invariant based on the pre-condition of the loop, the loop body, and the generated post-condition.
The loop invariant should satisfy: 1) it should be implied by the pre-condition of the loop; 2) it should be maintained after each iteration of the loop; 3) the conjunction of the loop invariant and the negation of the loop condition should imply the post-condition of the loop.
If \sys cannot find a valid loop invariant, the post-condition of the loop may be invalid.
\sys will generate another post-condition for the loop and repeat the above process.
Traditionally, developers usually first write a loop invariant and then use the loop invariant to derive the post-condition of the loop.
\sys adopts a different order, which first generates the post-condition of the loop and then tries to find a loop invariant to validate the post-condition.
This is because we observe that generating loop invariants is actually a more difficult task than directly generating the post-condition of the loop.
The post-condition just needs to hold after the loop, while the loop invariant needs to hold before and after each loop iteration.
\Cref{subsec:back:insight} also shows that LLMs can accurately predict the results of functions with loops.

In practice, reasoning about statements one by one may lead to numerous invocations of LLMs, which can be slow and expensive.
Thus, \sys does not infer the post-condition for every single statement.
It groups statements into larger blocks and reasons about them collectively, reducing the number of LLM invocations.
This works due to the insight that LLMs are able to accurately predict the execution results of small code blocks, which has been illustrated in \Cref{subsec:back:insight}.
For example, \sys can directly put the code block in \Cref{fig:reasoner:example} into the prompt and ask LLMs to infer the post-condition of the whole code block, which can save multiple invocations of LLMs for each statement in the code block.

Although \sys directly performs reasoning based on natural language pre-conditions and post-conditions, it can still be combined with formal methods to enhance the precision of the reasoning process.
Specifically, if a condition is precise enough, \sys can use LLMs to translate it into formulas.
Then, \sys can use an SMT solver to check the proof obligations, which can be more precise than reasoning based on natural language.
To check whether a natural language condition is precise enough, we observe that if a natural language condition is ambiguous, the generated formula will usually contain some uninterpreted functions, which can represent the ambiguous concepts in natural language.
Thus, if the generated formula contains some uninterpreted functions, \sys will not use the formula to check the proof obligations, but will still reason about the natural language conditions.

\section{Bug Validator}
\label{sec:validate}

To confirm the potential bugs and help understand the bug causes, \sys designs a bug validator.
As described in \Cref{sec:overview}, for each potential bug, the bug validator first leverages LLMs to generate a test case.
Then, it executes the test case on the system and checks the output.
The testing environment can be manually set up by developers in advance.
Note that \sys also supports developers writing markdown files to guide agents in setting up the testing environment.
Since the overall workflow of \sys is managed by agents, it is easy to add extra steps before executing \sys.
If the bug cannot be triggered after N attempts to generate test cases, the potential bug will be marked as unconfirmed and will not be reported as a genuine bug.
N can be set by developers, which is 10 in our evaluation.
This can help reduce false positives caused by LLM hallucinations.

The specific test case generation and bug confirmation strategies vary across systems, but \sys handles them automatically using LLMs based on the description of potential bugs.
For example, when checking bugs of a C compiler in our evaluation (\Cref{sec:eval}), the bug validator generates a C program as the test case, compiles it with the target compiler, and checks whether the expected bug is triggered.
Bugs such as compilation crashes can be directly observed, while incorrect compilation results are detected by compiling the same program with a reference compiler implementation (e.g., GCC) and comparing the outputs.
Similarly, when checking bugs of a SQL query engine, the bug validator executes the generated test cases and compares the results against a reference implementation (e.g., DuckDB).
For systems without a reference implementation, \sys instead checks execution results against the expected outcomes described in the specifications.

\section{Incremental Reasoning}
\label{sec:incremental}

Software usually evolves incrementally, which means that developers typically modify only a subset of the codebase at a time for bug fixes, feature additions, or requirement changes.
Re-reasoning over the entire codebase after each change is both prohibitively expensive for large systems and unnecessary.
\sys therefore employs incremental reasoning, which updates specifications and reasoning results of only some of the functions based on the previous run of \sys rather than restarting from scratch.

The core challenge is how to accurately identify functions that require updated specifications to reflect the developer goal, and which functions should be re-checked.
A natural approach is to limit reasoning to functions that are directly modified or newly added.
However, this is insufficient, because developers may inadvertently omit changes to functions that are necessary to realize the developer goal, and \sys would then miss the corresponding bugs.
Asking the LLM to identify such omitted functions is also unreliable, since the LLM itself can be the source of incomplete code changes during vibe coding.

To address this challenge, \sys uses the natural-language description of the developer goal to identify \emph{relevant functions}.
A function is relevant if its specification may need updates to remain consistent with the new goal.
The developer goal can be supplied manually or obtained automatically from git commit messages or vibe coding prompts.
\sys then tries to update specifications for relevant, modified, and newly added functions.
A function is subsequently re-reasoned about if it satisfies one of the following conditions: 
1) it is modified or newly added,
2) its specification is updated, and
3) it calls a function whose specification is updated.

\paragraph{Multi-grained relevant function identification.}
Given a developer goal, \sys employs a coarse-to-fine multi-grained search algorithm that narrows the scope of relevant functions across three granularities: modules, files, and functions.
The initial scope of relevant functions is the entire codebase, and \sys progressively narrows the scope to several modules, several files, and finally several functions that are likely relevant to the developer goal.
First, at the module granularity, \sys uses an LLM to decompose the codebase into modules, summarize the semantics of each module, and select those modules likely containing relevant functions based on the natural-language summaries.
Second, at the \emph{file} granularity, \sys uses an LLM to analyze the source files within each selected module and identify files likely containing relevant functions.
Finally, at the \emph{function} granularity, \sys applies a heuristic scoring algorithm to rank functions by relevance and selects the top-$N$ as relevant functions, where $N$ is a developer-configurable parameter.
The relevance score of a function is a weighted sum of:
1) text overlap between the function and the developer goal description,
2) relevance scores of the function's callers and callees, and
3) the relevance score of the class to which the function belongs.
If the highest score falls below a predefined threshold, or if the number of functions in a file exceeds a predefined threshold, \sys further invokes an LLM to re-rank functions.

\paragraph{Specification update.}
Following the top-down caller-before-callee order described in \Cref{sec:spec}, \sys updates specifications such that a caller's specification is updated before those of its callees.
The functions requiring specification updates include: 1) functions added or modified since the last \sys run, and 2) functions identified as relevant.
Specifications of removed functions are discarded.
For newly added functions, \sys generates specifications from expected specifications, the developer goal, and the function's source code, using a method similar to that in \Cref{sec:spec}.
For existing functions, \sys presents the current specification, expected specification, source code, and developer goal to the LLM and asks it to determine whether an update is needed and, if so, how to revise it.
When a function's specification changes, \sys also asks the LLM to revisit the expected specifications of its callees; if any expected specification changes, \sys propagates the update to the corresponding callee by updating the callee's specification.
Additionally, the expected specifications of a function's callers are also updated to reflect the change in the function's specification.
\section{Implementation}
\label{sec:impl}

We implement \sys in Python, and the implementation consists of three components: a specification generator, a code reasoner, and a bug validator.
Directly invoking LLMs will usually be limited by TPM (tokens per minute) and RPM (requests per minute), which limits the concurrency of \sys.
Thus, all three components invoke LLMs via OpenRouter~\cite{openrouter}, which is a proxy for LLM services and supports much higher concurrency.
For each function, the specification generator writes the implementation, specification, and expected specifications of each callee into a separate file.
The code reasoner takes such a file as input and reports potential bugs in JSON files.
The bug validator takes JSON files as input and generates test cases to validate potential bugs.
Different systems use different programming languages and test-case execution commands.
To easily generalize \sys to various systems, we write some markdown files to guide an agent called OpenCode~\cite{opencode} and integrate it with other code in \sys to execute the workflow.
For example, it will generate different parsers for various programming languages and autonomously run test cases on different systems.
\section{Evaluation}
\label{sec:eval}

The evaluation aims to answer the following questions:
1) Could \sys scale to real-world large-scale systems?
2) How many newly discovered bugs could \sys effectively detect in large-scale systems?
3) What is the time and token cost of running \sys?

\subsection{Experimental Setup}
\label{sec:eval:setup}

We run all experiments on a machine with 16 vCPUs and 32 GiB memory.
It uses a third-generation Intel Xeon Scalable processor and runs Ubuntu 24.04 Server 64-bit.
All three components of \sys, including specification generator, code reasoner, and bug validator, use Claude Sonnet 4.6~\cite{sonnet}.
The maximum number of attempts to generate test cases is set to 10.

\begin{table}[t]
\caption{The complexity of systems that \sys successfully reasons about.}
\centering
\begin{tabular}{llrr}
\toprule
\textbf{System} & \textbf{Type} & \textbf{LoC} & \textbf{\#Functions} \\
\midrule
\ccc        & Compiler & 143k & 4,957 \\
\vibetensor & ML System & 108k & 3,031 \\
\vibeos     & Operating System & 15k & 452 \\
\vibedb     & Database & 11k & 109 \\
\midrule
\textbf{Total} & & 277k & 8,549 \\
\bottomrule
\end{tabular}
\label{tab:eval:complexity}
\end{table}

\sys is evaluated on four large-scale systems shown in \Cref{tab:eval:complexity}.
We evaluate the scale of each system in terms of lines of code (LoC) and the number of functions.
The number of LoC is measured by cloc~\cite{cloc}, which counts the number of lines of code excluding blank lines and comments.
We only count the LoC of source code files, excluding test files and documentation files.
Specifically, the LoC of each system ranges from 11k to 143k.
The number of functions in each system ranges from 109 to 4,910.

Here is a brief introduction to each system.
\begin{itemize}
    \item \ccc (Claude's C Compiler)~\cite{ccc}. \ccc is a C compiler that was autonomously developed by Anthropic using Claude Code.
    \ccc supports various C standards and hardware architectures.
    It implements all compilation components from scratch, such as the preprocessor, optimizer, linker, and executable file generator.
    \item \vibetensor~\cite{vibetensor,xu2026vibetensor}. \vibetensor is a deep learning framework that was autonomously developed by NVIDIA using its coding agent. \vibetensor covers various components, such as Cuda runtime, core runtime, and GPU kernels.
    \item \vibeos~\cite{vibeos}. \vibeos is an operating system vibe-coded from scratch using Claude Code. It implements various OS components, such as process management, memory management, and file systems. It also contains GUI and some user space applications. Our evaluation focuses on the kernel part of \vibeos, whose LoC and number of functions are reported in \Cref{tab:eval:complexity}.
    \item \vibedb~\cite{vibedb,wehrstein2025bespokeolap}. \vibedb provides SQL query processing engines developed from scratch using GPT-5.2 Codex. It achieves an order-of-magnitude speedup over DuckDB. \vibedb also provides scripts to generate query engines. For fairness, we do not re-generate the engine using these scripts, but directly reuse the artifact provided by the developers.
\end{itemize}

These systems are selected for the following reasons.
First, they are large enough and automatically developed by coding agents, which makes it harder to manually write specifications.
Thus, they are suitable for evaluating the scalability and bug detection capability of \sys.
Second, they are developed by different coding agents in various programming languages, such as C++, Python, Cuda, and Rust.
They also cover different domains, such as compilers, deep learning frameworks, operating systems, and databases.
This allows us to evaluate the generality of \sys.

We do not compare \sys with existing verification tools, such as symbolic execution engines~\cite{klee,serval}, Verus~\cite{verus,verus2024sosp}, and Dafny~\cite{dafny}.
This is because they usually require manually written accurate specifications, which is infeasible for large-scale systems in our evaluation.
Specifications generated by \sys cannot be used by these tools, as they target formal specifications rather than natural language specifications.
Note that these systems have been tested by their developers using various methods, which will be introduced in \Cref{subsec:eval:bug}.

\subsection{Capability of Bug Discovery}
\label{subsec:eval:bug}

\begin{table}[t]
\caption{The number of bugs newly discovered by \sys for each system.}
\centering
\begin{tabular}{lrr}
\toprule
\textbf{System} & \textbf{\#Bugs}  \\
\midrule
\ccc        &  339 \\
\vibetensor &  141 \\
\vibeos     &  23 \\
\vibedb     &  19 \\
\midrule
\textbf{Total} &  522 \\
\bottomrule
\end{tabular}
\label{tab:eval:bug}
\end{table}


\begin{table}[t]
\caption{The bug discovery methods already used by developers of each system.}
\centering
\begin{tabular}{lp{0.7\linewidth}}
\toprule
\textbf{System} & \textbf{Method} \\
\midrule
\ccc        &  Integration tests and unit tests \\
\vibetensor &  GoogleTest suites (208 cases), Pytest suites (203 cases), differential check against PyTorch, and multi-agent code review\\
\vibedb     &  Differential check against DuckDB \\
\bottomrule
\end{tabular}
\label{tab:eval:test}
\end{table}

To evaluate the bug discovery capability of \sys, we download the latest version (at submission time) of each system from GitHub repositories and run \sys on them.
Note that 3 of 4 systems have already been well tested and fixed by their developers (\Cref{tab:eval:test}).
\vibeos has been partially tested, but its developers do not release the testing details.
\Cref{tab:eval:bug} shows that \sys can still discover 522 new bugs.
The evaluation results demonstrate the effectiveness of \sys in discovering previously undetected bugs.

We further analyze the contribution of our techniques to bug discovery.
For example, we implement and run an ablation version on the largest system \ccc in our evaluation.
The ablation version generates specifications only based on function implementations rather than using our top-down paradigm.
Besides, it does not use our code reasoner to check the implementation against the specification.
Instead, it directly asks LLMs to answer whether the implementation satisfies its specification without Hoare-style reasoning.
Finally, it uses the same bug validator as \sys to confirm potential bugs.
As a result, it discovers only 57 bugs in \ccc, which is much fewer than 339 bugs discovered by \sys.

We further analyze the bugs reported by \sys to show that they can cause serious consequences.
In \ccc, \sys discovers 339 bugs.
Here are some of the main bug types.
111 bugs cause incorrect code generation at the IR level.
68 bugs involve incorrect runtime outputs.
39 bugs result in missing or incorrect diagnostic messages.
14 bugs result in compilation crash or hang.
9 bugs are about compilation optimization, such as failing to remove unnecessary instructions.


In \vibeos, \sys discovers 23 bugs.
Specifically, 4 bugs cause memory corruption or stack overflow.
4 bugs result in incorrect return values of system calls.
2 bugs lead to infinite loops.
1 bug causes a failed process creation to still increment the PID counter.
The remaining 12 bugs involve incorrect or missing logic in code.
For example, net_ping always fails immediately if the timeout value is under 10ms, because integer division truncates the loop bound to zero.


In \vibedb, \sys discovers 19 bugs.
Specifically, 7 bugs cause the query execution to return incorrect results.
\vibedb synthesizes SQL query engines for specific SQL queries, such as the TPC-H benchmark.
It generates C++ functions to execute the workflow of SQL queries, which brings significant performance improvement.
However, some of these functions contain bugs and break the intended semantics.
For example, one bug causes a number in the query result to be truncated rather than rounded.
3 bugs lead to process crashes.
9 bugs result in silent acceptance of invalid inputs.


In \vibetensor, \sys discovers 141 bugs.
Here are some of the main bug types.
Specifically, 50 bugs cause silently incorrect execution results, such as incorrect tensor values, shapes, or data types.
51 bugs result in missing or incorrect error handling.
19 bugs cause memory safety issues, such as memory leaks.
3 bugs lead to program crashes.


\subsection{Scalability of \sys} 
\label{subsec:eval:scalability}

Compositional reasoning of systems in \Cref{tab:eval:complexity} takes about 2 days and 3.4 billion tokens in total.
There is no existing verification tool that can handle all these large-scale systems.
Although \sys cannot ensure soundness, it realizes much better scalability than existing verification tools.
However, note that we do not claim \sys can replace existing formal verification tools.
\sys realizes high scalability, while prior verification tools ensure soundness.
They target different scenarios.

\begin{table}[t]
\caption{The concurrency space of specification generation for each system, including the number of phases and the maximum/median/minimum numbers of layers in each phase.
More phases and fewer layers per phase indicate higher concurrency in specification generation.}
\centering
\begin{tabular}{lrrrr}
\toprule
\multirow{2}{*}{\textbf{System}} & \multirow{2}{*}{\textbf{\#Phases}} & \multicolumn{3}{c}{\textbf{\#Layers per Phase}} \\
\cmidrule(l){3-5}
 &  & \textbf{Max} & \textbf{Median} & \textbf{Min}  \\
\midrule
\ccc        & 13 & 19 & 10 & 4 \\
\vibetensor & 12 & 18 & 7 & 2 \\
\vibeos     & 14 & 9 & 4 & 1 \\
\vibedb     &  6 & 4 & 3 & 1 \\
\bottomrule
\end{tabular}
\label{tab:eval:concurrency}
\end{table}

The scalability of \sys can be attributed to its top-down specification generation paradigm, which allows concurrency for the specification generator, code reasoner, and bug validator.
This is critical for scaling to large systems, as LLM invocation is time-consuming.
Regarding the specification generator, \sys decomposes the system into self-contained phases and splits each phase into layers based on function call graphs.
As shown in \Cref{tab:eval:concurrency}, the number of phases ranges from 6 to 14, and the number of layers per phase ranges from 1 to 19.
The specification generation of different phases can be executed concurrently. Within each phase, the specification generation of functions in the same layer can also be executed concurrently.
Although the layers in each system should be handled sequentially in top-down order, the maximum number of layers per phase is 19.
This still provides concurrency space.
For code reasoner and bug validator, \sys can handle each function independently and concurrently, which can further improve the scalability of \sys.
This comes from the function-level specification, which enables compositional reasoning in \sys.

\section{Related Work}
\label{sec:related}

\paragraph{Formal verification.}
Recent work has made strong progress in formal verification for systems.
Some works~\cite{fscq,dfscq,atomfs,certikos,Klewin2009sel4,sosp23grove,osdi23vmvcc,tej2021gojournal,perennial,cspec,Tao2021VRM,Li2022realms,xupeng2023spoq,zou2024reffs} use interactive theorem provers, such as Coq~\cite{coq}, to verify system correctness against formal specifications.
They introduce new theories and verify critical properties.
However, they require manual proofs.
This creates a heavy human burden and does not scale well to large systems.
For example, FSCQ~\cite{fscq}, a sequential verified file system, takes several researchers about 1.5 years to complete.
Other works~\cite{chen2025atmosphere,zhang2025automan,sun2024anvil,zhou2024verismo,ironsync,armada,veribetrkv,komodo,ironfleet} use semi-automated verifiers to reduce manual proof effort.
However, developers still need to provide formal specifications, loop invariants, and partial proofs.
This also limits scalability.
For example, implementing and verifying a cluster management controller with Anvil~\cite{sun2024anvil} takes around 2.5 person-months, with a proof-to-code ratio ranging from 4.5 to 7.4 across different controllers.
Other works based on symbolic execution~\cite{Nelson2017hyperkernel,yv6,serval,Nelson2020bpf,klee,Helgi2018nickel,sosp2019vigor} achieve fully automated verification.
However, they struggle with path explosion and cannot handle unbounded loops, which are common in system code.
\sys avoids manual effort and scales to large systems.
Although it does not ensure soundness, our evaluation shows that it effectively finds many bugs in real-world systems.
Therefore, \sys offers a feasible direction for improving system reliability, especially in the era of LLMs.
Note that \sys does not aim to replace existing formal verifiers.
Rather, it aims to complement existing works by providing a practical method to reason about large systems and find bugs.
Integrating existing verification theories with \sys is an interesting future direction that can further improve system reliability.

\paragraph{LLM-assisted formal verification.}
To reduce human proof effort, some works~\cite{yang2025autoverus,qin2025hotos,mugnier2025laurel,chen2024safe} combine LLMs with traditional verification techniques.
The key idea is to use LLMs to automatically generate proofs or annotations (e.g., loop invariants and assertions) for verification tools (e.g., Coq and Verus), enabling fully automated verification.
Although these works reduce human proof effort, they focus on generating proofs rather than specifications.
They still require pre-defined formal specifications, which is a major barrier to compositional reasoning for large systems.
Moreover, LLM-generated code makes manually writing specifications harder, as developers lack a deep understanding of the code.
In contrast, \sys automatically generates specifications, enabling automated compositional reasoning.

\paragraph{Generation of specifications.}
Prior works~\cite{chen2024safe,lahirie2024evaluating} have explored automatically generating specifications for programs. 
For example, SAFE leverages LLMs to synthesize formal specifications for Rust programs.
Its specification generation heavily relies on analyzing the program implementation and reflects implementation behavior.
This is because SAFE generates specifications to synthesize training data, where each sample is a Rust program paired with a formal specification and proof.
The training data are then used to enhance the capability of LLMs to generate formal proofs.
As a result, SAFE and \sys target different goals, and \sys cannot reuse the method of SAFE.
In \sys, the generated specifications guide the reasoning process.
They reflect the expected behavior of functions from the system's perspective, regardless of implementation bugs.
Some other works, such as \sysspec~\cite{liu2026sharpen}, propose a new paradigm for manually writing specifications that can guide LLMs to automatically generate file systems.
Unlike \sysspec, \sys focuses on automatically generating specifications for existing systems, which helps reason about system correctness.
\sysspec relies on standard test suites rather than reasoning to check correctness.

\section{Discussion}
\label{sec:discussion}

\paragraph{\sys vs. formal verifiers}
There is no automated formal verifier that can scale to systems in our evaluation, but \sys does not aim to replace them.
\sys and formal verifiers target different scenarios and have their own advantages.
Formal verifiers ensure soundness, while \sys targets scalable reasoning.
How to combine \sys with formal verifiers to achieve better results is an interesting future research direction.

\paragraph{Reasoning vs. testing.}
Both testing and reasoning are important for improving large-scale system reliability.
Testing is lightweight and scales well, while reasoning analyzes code semantics to find bugs that testing may miss.
The evaluation shows \sys finds bugs that testing methods cannot detect.
\sys enables reasoning about large-scale systems and can be combined with testing to improve reliability.
Additionally, \sys-generated specifications can guide testing.
Testing requires a testing oracle to verify results, which is difficult to construct automatically.
Using \sys specifications to help build testing oracles is a promising research direction.

\paragraph{Support for concurrency.}
Currently, \sys focuses on reasoning about sequential programs.
Since we have integrated Hoare logic inference rules into \sys to reason about sequential programs, it may also be feasible to integrate more theories about concurrent programs, such as rely-guarantee~\cite{Liang2013mod,rgsim} and concurrent separation logic~\cite{ralf2015iris,OHearn2007csl}.
It is a possible future direction to extend \sys to support reasoning about concurrent programs.

\paragraph{Reasoning about LLM-based applications.}
LLMs are increasingly integrated into various applications, creating new challenges for reasoning about their behavior.
It is difficult to formally define what properties LLM outputs must satisfy.
One possible way is to assume LLM outputs satisfy the requirements in prompts, usually expressed as natural language rather than formulas.
This scenario requires reasoning about code against natural language specifications like \sys.
Extending \sys to reason about LLM-based applications is a promising future research direction.

\section{Conclusion}
\label{sec:conclusion}

This paper presents the first framework that enables automated compositional reasoning for large-scale systems.
The evaluation shows that \sys can successfully find newly discovered bugs in systems with up to 143k LoC per system.
The bugs can cause serious consequences, including system crashes and incorrect execution results.

\bibliographystyle{ACM-Reference-Format}
\bibliography{paper}

@article{hoarelogic,
author = {Hoare, C. A. R.},
title = {An axiomatic basis for computer programming},
year = {1969},
issue_date = {Oct. 1969},
publisher = {Association for Computing Machinery},
address = {New York, NY, USA},
volume = {12},
number = {10},
issn = {0001-0782},
url = {https://doi.org/10.1145/363235.363259},
doi = {10.1145/363235.363259},
journal = {Commun. ACM},
month = oct,
pages = {576–580},
numpages = {5}
}

@article{mugnier2025laurel,
author = {Mugnier, Eric and Gonzalez, Emmanuel Anaya and Polikarpova, Nadia and Jhala, Ranjit and Yuanyuan, Zhou},
title = {Laurel: Unblocking Automated Verification with Large Language Models},
year = {2025},
issue_date = {April 2025},
publisher = {Association for Computing Machinery},
address = {New York, NY, USA},
volume = {9},
number = {OOPSLA1},
url = {https://doi.org/10.1145/3720499},
doi = {10.1145/3720499},
journal = {Proc. ACM Program. Lang.},
month = apr,
articleno = {134},
numpages = {27}
}

@inproceedings{qin2025hotos,
author = {Qin, Jianxing and Du, Alexander and Zhang, Danfeng and Lentz, Matthew and Zhuo, Danyang},
title = {Can Large Language Models Verify System Software? A Case Study Using FSCQ as a Benchmark},
year = {2025},
isbn = {9798400714757},
publisher = {Association for Computing Machinery},
address = {New York, NY, USA},
url = {https://doi.org/10.1145/3713082.3730382},
doi = {10.1145/3713082.3730382},
booktitle = {Proceedings of the 2025 Workshop on Hot Topics in Operating Systems},
pages = {34–41},
numpages = {8},
location = {Banff, AB, Canada},
series = {HotOS '25}
}

@inproceedings{sosp2019vigor,
author = {Zaostrovnykh, Arseniy and Pirelli, Solal and Iyer, Rishabh and Rizzo, Matteo and Pedrosa, Luis and Argyraki, Katerina and Candea, George},
title = {Verifying software network functions with no verification expertise},
year = {2019},
isbn = {9781450368735},
publisher = {Association for Computing Machinery},
address = {New York, NY, USA},
url = {https://doi.org/10.1145/3341301.3359647},
doi = {10.1145/3341301.3359647},
booktitle = {Proceedings of the 27th ACM Symposium on Operating Systems Principles},
pages = {275–290},
numpages = {16},
location = {Huntsville, Ontario, Canada},
series = {SOSP '19}
}

@inproceedings{komodo,
  author    = {Ferraiuolo, Andrew and Baumann, Andrew and Hawblitzel, Chris and Parno, Bryan},
  booktitle = {Proceedings of the ACM Symposium on Operating Systems Principles (SOSP)},
  code      = {https://github.com/Microsoft/Komodo},
  month     = {October},
  title     = {Komodo: Using verification to disentangle secure-enclave hardware from software},
  year      = {2017}
}

@inproceedings{veribetrkv,
  author    = {Hance, Travis and Lattuada, Andrea and Hawblitzel, Chris and Howell, Jon and Johnson, Rob and Parno, Bryan},
  booktitle = {Proceedings of the USENIX Symposium on Operating Systems Design and Implementation (OSDI)},
  code      = {https://github.com/secure-foundations/veribetrkv-osdi2020},
  month     = {November},
  title     = {Storage Systems are Distributed Systems (So Verify Them That Way!)},
  year      = {2020}
}

@inproceedings{ironsync,
  author    = {Hance, Travis and Zhou, Yi and Lattuada, Andrea and Achermann, Reto and Conway, Alex and Stutsman, Ryan and Zellweger, Gerd and Hawblitzel, Chris and Howell, Jon and Parno, Bryan},
  booktitle = {Proceedings of the USENIX Symposium on Operating Systems Design and Implementation (OSDI)},
  code      = {https://github.com/secure-foundations/ironsync-osdi2023},
  month     = {July},
  title     = {Sharding the State Machine: Automated Modular Reasoning for Complex Concurrent Systems},
  year      = {2023}
}

@inproceedings {zhou2024verismo,
author = {Ziqiao Zhou and Anjali and Weiteng Chen and Sishuai Gong and Chris Hawblitzel and Weidong Cui},
title = {{VeriSMo}: A Verified Security Module for Confidential {VMs}},
booktitle = {18th USENIX Symposium on Operating Systems Design and Implementation (OSDI 24)},
year = {2024},
isbn = {978-1-939133-40-3},
address = {Santa Clara, CA},
pages = {599--614},
url = {https://www.usenix.org/conference/osdi24/presentation/zhou},
publisher = {USENIX Association},
month = jul
}

@inproceedings {sun2024anvil,
author = {Xudong Sun and Wenjie Ma and Jiawei Tyler Gu and Zicheng Ma and Tej Chajed and Jon Howell and Andrea Lattuada and Oded Padon and Lalith Suresh and Adriana Szekeres and Tianyin Xu},
title = {Anvil: Verifying Liveness of Cluster Management Controllers},
booktitle = {18th USENIX Symposium on Operating Systems Design and Implementation (OSDI 24)},
year = {2024},
isbn = {978-1-939133-40-3},
address = {Santa Clara, CA},
pages = {649--666},
url = {https://www.usenix.org/conference/osdi24/presentation/sun-xudong},
publisher = {USENIX Association},
month = jul
}

@inproceedings {zou2024reffs,
author = {Mo Zou and Dong Du and Mingkai Dong and Haibo Chen},
title = {Using Dynamically Layered Definite Releases for Verifying the {RefFS} File System},
booktitle = {18th USENIX Symposium on Operating Systems Design and Implementation (OSDI 24)},
year = {2024},
isbn = {978-1-939133-40-3},
address = {Santa Clara, CA},
pages = {629--648},
url = {https://www.usenix.org/conference/osdi24/presentation/zou},
publisher = {USENIX Association},
month = jul
}

@inproceedings{zhang2025automan,
author = {Zhang, Zihao and Zhou, Ti and Jenkins, Christa and Chowdhury, Omar and Mu, Shuai},
title = {AutoMan: Facilitating Verified Distributed Systems Development Through Automatic Code Generation and Manual Optimizations},
year = {2025},
isbn = {9798400718700},
publisher = {Association for Computing Machinery},
address = {New York, NY, USA},
url = {https://doi.org/10.1145/3731569.3764822},
doi = {10.1145/3731569.3764822},
booktitle = {Proceedings of the ACM SIGOPS 31st Symposium on Operating Systems Principles},
pages = {768–785},
numpages = {18},
location = {Lotte Hotel World, Seoul, Republic of Korea},
series = {SOSP '25}
}

@inproceedings{chen2025atmosphere,
author = {Chen, Xiangdong and Li, Zhaofeng and Zhang, Jerry and Narayanan, Vikram and Burtsev, Anton},
title = {Atmosphere: Practical Verified Kernels with Rust and Verus},
year = {2025},
isbn = {9798400718700},
publisher = {Association for Computing Machinery},
address = {New York, NY, USA},
url = {https://doi.org/10.1145/3731569.3764821},
doi = {10.1145/3731569.3764821},
booktitle = {Proceedings of the ACM SIGOPS 31st Symposium on Operating Systems Principles},
pages = {752–767},
numpages = {16},
location = {Lotte Hotel World, Seoul, Republic of Korea},
series = {SOSP '25}
}

@inproceedings {xupeng2023spoq,
author = {Xupeng Li and Xuheng Li and Wei Qiang and Ronghui Gu and Jason Nieh},
title = {Spoq: Scaling {Machine-Checkable} Systems Verification in Coq},
booktitle = {17th USENIX Symposium on Operating Systems Design and Implementation (OSDI 23)},
year = {2023},
isbn = {978-1-939133-34-2},
address = {Boston, MA},
pages = {851--869},
url = {https://www.usenix.org/conference/osdi23/presentation/li-xupeng},
publisher = {USENIX Association},
month = jul
}

@inproceedings{osdi23vmvcc,
  title        = {Verifying {vMVCC}, a high-performance transaction
    library using multi-version concurrency control},
  author       = {Yun-Sheng Chang and Ralf Jung and Upamanyu Sharma
    and Joseph Tassarotti and M. Frans Kaashoek and Nickolai Zeldovich},
  booktitle    = {Proceedings of the 17th {USENIX} {S}ymposium on
    {O}perating {S}ystems {D}esign and {I}mplementation ({OSDI} '23)},
  year         = 2023,
  month        = jul,
}

@inproceedings{sosp23grove,
  title        = {{G}rove: a {S}eparation-{L}ogic {L}ibrary for
    {V}erifying {D}istributed {S}ystems},
  author       = {Upamanyu Sharma and Ralf Jung and Joseph Tassarotti
    and Frans Kaashoek and Nickolai Zeldovich},
  pages        = {113--129},
  booktitle    = {Proceedings of the 29th ACM Symposium on Operating
    Systems Principles (SOSP 2023)},
  year         = 2023,
  month        = oct,
  address      = {Koblenz, Germany},
}

@inproceedings {Li2022realms,
author = {Xupeng Li and Xuheng Li and Christoffer Dall and Ronghui Gu and Jason Nieh and Yousuf Sait and Gareth Stockwell},
title = {Design and {V}erification of the {A}rm {C}onfidential {C}ompute {A}rchitecture},
booktitle = {16th USENIX Symposium on Operating Systems Design and Implementation},
year = {2022},
isbn = {978-1-939133-28-1},
address = {Carlsbad, CA},
pages = {465--484},
url = {https://www.usenix.org/conference/osdi22/presentation/li},
publisher = {USENIX Association},
month = jul,
}

@article{Liang2013mod,
author = {Liang, Hongjin and Feng, Xinyu},
title = {Modular {V}erification of {L}inearizability with {N}on-{F}ixed {L}inearization {P}oints},
year = {2013},
issue_date = {June 2013},
publisher = {Association for Computing Machinery},
address = {New York, NY, USA},
volume = {48},
number = {6},
issn = {0362-1340},
url = {https://doi.org/10.1145/2499370.2462189},
doi = {10.1145/2499370.2462189},
journal = {SIGPLAN Not.},
month = {jun},
pages = {459-470},
numpages = {12}
}

@inproceedings{fscq,
author = {Chen, Haogang and Ziegler, Daniel and Chajed, Tej and Chlipala, Adam and Kaashoek, M. Frans and Zeldovich, Nickolai},
title = {Using {C}rash {H}oare {L}ogic for {C}ertifying the {FSCQ} {F}ile {S}ystem},
year = {2015},
isbn = {9781450338349},
publisher = {Association for Computing Machinery},
address = {New York, NY, USA},
booktitle = {Proceedings of the 25th Symposium on Operating Systems Principles},
pages = {18--37},
numpages = {20},
location = {Monterey, California},

}

@inproceedings {yv6,
author = {Helgi Sigurbjarnarson and James Bornholt and Emina Torlak and Xi Wang},
title = {Push-{B}utton {V}erification of {F}ile {S}ystems via {C}rash {R}efinement},
booktitle = {12th USENIX Symposium on Operating Systems Design and Implementation},
year = {2016},
isbn = {978-1-931971-33-1},
address = {Savannah, GA},
pages = {1--16},
url = {https://www.usenix.org/conference/osdi16/technical-sessions/presentation/sigurbjarnarson},
publisher = {{USENIX} Association},
month = nov,
}

@inproceedings{dfscq,
author = {Chen, Haogang and Chajed, Tej and Konradi, Alex and Wang, Stephanie and undefinedleri, Atalay and Chlipala, Adam and Kaashoek, M. Frans and Zeldovich, Nickolai},
title = {Verifying a {H}igh-{P}erformance {C}rash-{S}afe {F}ile {S}ystem {U}sing a {T}ree {S}pecification},
year = {2017},
isbn = {9781450350853},
publisher = {Association for Computing Machinery},
address = {New York, NY, USA},
booktitle = {Proceedings of the 26th Symposium on Operating Systems Principles},
pages = {270--286},
numpages = {17},
location = {Shanghai, China}
}

@inproceedings{armada,
author = {Lorch, Jacob R. and Chen, Yixuan and Kapritsos, Manos and Parno, Bryan and Qadeer, Shaz and Sharma, Upamanyu and Wilcox, James R. and Zhao, Xueyuan},
title = {Armada: {L}ow-{E}ffort {V}erification of {H}igh-{P}erformance {C}oncurrent {P}rograms},
year = {2020},
isbn = {9781450376136},
publisher = {Association for Computing Machinery},
address = {New York, NY, USA},
booktitle = {Proceedings of the 41st ACM SIGPLAN Conference on Programming Language Design and Implementation},
pages = {197--210},
numpages = {14},
location = {London, UK},

}

@inproceedings {cspec,
author = {Tej Chajed and Frans Kaashoek and Butler Lampson and Nickolai Zeldovich},
title = {Verifying {C}oncurrent {S}oftware {U}sing {M}overs in {CSPEC}},
booktitle = {13th {USENIX} Symposium on Operating Systems Design and Implementation},
year = {2018},
isbn = {978-1-939133-08-3},
address = {Carlsbad, CA},
pages = {306--322},
url = {https://www.usenix.org/conference/osdi18/presentation/chajed},
publisher = {{USENIX} Association},
month = oct,
}

@inproceedings {certikos,
author = {Ronghui Gu and Zhong Shao and Hao Chen and Xiongnan (Newman) Wu and Jieung Kim and Vilhelm Sj{\"o}berg and David Costanzo},
title = {CertiKOS: {A}n {E}xtensible {A}rchitecture for {B}uilding {C}ertified {C}oncurrent {OS} {K}ernels},
booktitle = {12th {USENIX} Symposium on Operating Systems Design and Implementation},
year = {2016},
isbn = {978-1-931971-33-1},
address = {Savannah, GA},
pages = {653--669},
url = {https://www.usenix.org/conference/osdi16/technical-sessions/presentation/gu},
publisher = {{USENIX} Association},
month = nov,
}

@inproceedings{serval,
author = {Nelson, Luke and Bornholt, James and Gu, Ronghui and Baumann, Andrew and Torlak, Emina and Wang, Xi},
title = {Scaling {S}ymbolic {E}valuation for {A}utomated {V}erification of {S}ystems {C}ode with {S}erval},
year = {2019},
publisher = {Association for Computing Machinery},
address = {New York, NY, USA},
booktitle = {Proceedings of the 27th ACM Symposium on Operating Systems Principles},
pages = {225--242},
numpages = {18},
location = {Huntsville, Ontario, Canada},

}

@inproceedings{rgsim,
author = {Liang, Hongjin and Feng, Xinyu and Fu, Ming},
title = {A {R}ely-{G}uarantee-{B}ased {S}imulation for {V}erifying {C}oncurrent {P}rogram {T}ransformations},
year = {2012},
publisher = {Association for Computing Machinery},
address = {New York, NY, USA},
booktitle = {Proceedings of the 39th Annual ACM SIGPLAN-SIGACT Symposium on Principles of Programming Languages},
pages = {455--468},
numpages = {14},
location = {Philadelphia, PA, USA},

}

@inproceedings{ironfleet,
author = {Hawblitzel, Chris and Howell, Jon and Kapritsos, Manos and Lorch, Jacob R. and Parno, Bryan and Roberts, Michael L. and Setty, Srinath and Zill, Brian},
title = {Iron{F}leet: {P}roving {P}ractical {D}istributed {S}ystems {C}orrect},
year = {2015},
isbn = {9781450338349},
publisher = {Association for Computing Machinery},
address = {New York, NY, USA},
url = {https://doi.org/10.1145/2815400.2815428},
doi = {10.1145/2815400.2815428},
booktitle = {Proceedings of the 25th Symposium on Operating Systems Principles},
pages = {1--17},
numpages = {17},
location = {Monterey, California},

}

@inproceedings{perennial,
  title        = {Verifying {C}oncurrent, {C}rash-{S}afe {S}ystems with
    {{P}erennial}},
  author       = {Tej Chajed and Joseph Tassarotti and M. Frans
    Kaashoek and Nickolai Zeldovich},
  booktitle    = {Proceedings of the 27th ACM Symposium on Operating
    Systems Principles},
  year         = 2019,
  month        = oct,
  address      = {Hunstville, ON, Canada},
}

@inproceedings{atomfs,
author = {Zou, Mo and Ding, Haoran and Du, Dong and Fu, Ming and Gu, Ronghui and Chen, Haibo},
title = {Using {C}oncurrent {R}elational {L}ogic with {H}elpers for {V}erifying the {A}tom{FS} {F}ile {S}ystem},
year = {2019},
isbn = {9781450368735},
publisher = {Association for Computing Machinery},
address = {New York, NY, USA},
booktitle = {Proceedings of the 27th ACM Symposium on Operating Systems Principles},
pages = {259-274},
numpages = {16},
location = {Huntsville, Ontario, Canada},

}

@inproceedings {tej2021gojournal,
author = {Tej Chajed and Joseph Tassarotti and Mark Theng and Ralf Jung and M. Frans Kaashoek and Nickolai Zeldovich},
title = {Go{J}ournal: {A} {V}erified, {C}oncurrent, {C}rash-safe {J}ournaling {S}ystem},
booktitle = {15th {USENIX} Symposium on Operating Systems Design and Implementation},
year = {2021},
isbn = {978-1-939133-22-9},
pages = {423--439},
url = {https://www.usenix.org/conference/osdi21/presentation/chajed},
publisher = {{USENIX} Association},
month = jul,
}

@inproceedings{Tao2021VRM,
author = {Tao, Runzhou and Yao, Jianan and Li, Xupeng and Li, Shih-Wei and Nieh, Jason and Gu, Ronghui},
title = {Formal {V}erification of a {M}ultiprocessor {H}ypervisor on {A}rm {R}elaxed {M}emory {H}ardware},
year = {2021},
isbn = {9781450387095},
publisher = {Association for Computing Machinery},
address = {New York, NY, USA},
url = {https://doi.org/10.1145/3477132.3483560},
doi = {10.1145/3477132.3483560},
booktitle = {Proceedings of the ACM SIGOPS 28th Symposium on Operating Systems Principles},
pages = {866-881},
numpages = {16},
location = {Virtual Event, Germany},

}

@inproceedings{ralf2015iris,
author = {Ralf, Jung and David, Swasey and Filip, Sieczkowski and Kasper, Svendsen and Aaron, Turon and Lars, Birkedal and Derek, Dreyer},
title = {Iris: {M}onoids and {I}nvariants as an {O}rthogonal {B}asis for {C}oncurrent {R}easoning},
year = {2015},
isbn = {9781450333009},
publisher = {Association for Computing Machinery},
address = {New York, NY, USA},
url = {https://doi.org/10.1145/2676726.2676980},
doi = {10.1145/2676726.2676980},
booktitle = {Proceedings of the 42nd Annual ACM SIGPLAN-SIGACT Symposium on Principles of Programming Languages},
pages = {637-650},
numpages = {14},
location = {Mumbai, India},

}

@article{Klewin2009sel4,
author = {Klein, Gerwin},
year = {2009},
month = {Feb},
pages = {27-69},
title = {Operating {S}ystem {V}erification—{A}n {O}verview},
volume = {34},
journal = {Sadhana},
doi = {10.1007/s12046-009-0002-4}
}

@article{OHearn2007csl,
author = {OHearn, Peter W.},
title = {Resources, {C}oncurrency, and {L}ocal {R}easoning},
year = {2007},
issue_date = {April, 2007},
publisher = {Elsevier Science Publishers Ltd.},
address = {GBR},
volume = {375},
number = {1-3},
issn = {0304-3975},
url = {https://doi.org/10.1016/j.tcs.2006.12.035},
doi = {10.1016/j.tcs.2006.12.035},
journal = {Theor. Comput. Sci.},
month = {apr},
pages = {271-307},
numpages = {37}
}

@inproceedings{Nelson2017hyperkernel,
author = {Nelson, Luke and Sigurbjarnarson, Helgi and Zhang, Kaiyuan and Johnson, Dylan and Bornholt, James and Torlak, Emina and Wang, Xi},
title = {Hyperkernel: {P}ush-{B}utton {V}erification of an OS {K}ernel},
year = {2017},
isbn = {9781450350853},
publisher = {Association for Computing Machinery},
address = {New York, NY, USA},
url = {https://doi.org/10.1145/3132747.3132748},
doi = {10.1145/3132747.3132748},
booktitle = {Proceedings of the 26th Symposium on Operating Systems Principles},
pages = {252-269},
numpages = {18},
location = {Shanghai, China},

}

@inproceedings {Helgi2018nickel,
author = {Helgi Sigurbjarnarson and Luke Nelson and Bruno Castro-Karney and James Bornholt and Emina Torlak and Xi Wang},
title = {Nickel: A {F}ramework for {D}esign and {V}erification of {I}nformation {F}low {C}ontrol {S}ystems},
booktitle = {13th USENIX Symposium on Operating Systems Design and Implementation},
year = {2018},
isbn = {978-1-939133-08-3},
address = {Carlsbad, CA},
pages = {287--305},
url = {http://www.usenix.org/conference/osdi18/presentation/sigurbjarnarson},
publisher = {USENIX Association},
month = oct,
}

@inbook{Nelson2020bpf,
author = {Nelson, Luke and Van Geffen, Jacob and Torlak, Emina and Wang, Xi},
title = {Specification and {V}erification in the {F}ield: {A}pplying {F}ormal {M}ethods to BPF {J}ust-in-{T}ime {C}ompilers in the {L}inux {K}ernel},
year = {2020},
isbn = {978-1-939133-19-9},
publisher = {USENIX Association},
address = {USA},
booktitle = {Proceedings of the 14th USENIX Conference on Operating Systems Design and Implementation},
articleno = {3},
numpages = {21}
}

@misc{verusbench,
  author       = {Microsoft},
  year = {2026},
  title        = {VerusBench},
  howpublished = {\url{https://github.com/microsoft/verus-proof-synthesis/tree/main/benchmarks/VerusBench},},
}

@article{wehrstein2025bespokeolap,
  title     = {Bespoke OLAP: Synthesizing Workload-Specific One-Size-Fits-One Database Engines},
  author    = {Wehrstein, Johannes and Eckmann, Timo and Jasny, Matthias and Binnig, Carsten},
  journal   = {arXiv preprint arXiv:2603.02001},
  year      = {2025},
  url       = {https://arxiv.org/abs/2603.02001}
}

@misc{vibedb,
  author       = {Wehrstein, Johannes and Eckmann, Timo and Jasny, Matthias and Binnig, Carsten},
  year = {2026},
  title        = {Bespoke-OLAP},
  howpublished = {\url{https://github.com/DataManagementLab/BespokeOLAP},},
}

@misc{vibeos,
  author       = {Kaanse},
  year = {2026},
  title        = {VibeOS},
  howpublished = {\url{https://github.com/kaansenol5/VibeOS/tree/main},},
}

@misc{ccc,
  author       = {Anthropics},
  year = {2026},
  title        = {CCC — Claude's C Compiler},
  howpublished = {\url{https://github.com/anthropics/claudes-c-compiler/},},
}

@misc{xu2026vibetensor,
Author = {Bing Xu and Terry Chen and Fengzhe Zhou and Tianqi Chen and Yangqing Jia and Vinod Grover and Haicheng Wu and Wei Liu and Craig Wittenbrink and Wen-mei Hwu and Roger Bringmann and Ming-Yu Liu and Luis Ceze and Michael Lightstone and Humphrey Shi},
Title = {VibeTensor: System Software for Deep Learning, Fully Generated by AI Agents},
Year = {2026},
Eprint = {arXiv:2601.16238},
}

@misc{sonnet,
  author       = {Anthropic},
  title        = {Claude Sonnet 4.6},
  year = {2026},
  howpublished = {\url{https://www.anthropic.com/claude/sonnet},},
}

@misc{openaicodex,
  author       = {OpenAI},
  year = {2026},
  title        = {OpenAI Codex},
  howpublished = {\url{https://openai.com/codex/},},
}

@misc{githubcopilot,
  author       = {GitHub},
  year = {2026},
  title        = {GitHub Copilot},
  howpublished = {\url{https://github.com/features/copilot},},
}

@misc{cursor,
  author       = {Anysphere},
  year = {2026},
  title        = {Cursor},
  howpublished = {\url{https://cursor.com/},},
}

@misc{claudecode,
  author       = {Anthropic},
  year = {2026},
  title        = {Claude Code},
  howpublished = {\url{https://claude.com/product/claude-code},},
}

@misc{vibetensor,
  author       = {NVIDIA},
  year = {2026},
  title        = {VibeTensor},
  howpublished = {\url{https://github.com/NVlabs/vibetensor},},
}

@misc{coq,
  author       = {Coq Development Team},
  title        = {The {C}oq {P}roof {A}ssistant},
  year         = {2024},
  howpublished = {\url{https://coq.inria.fr/},},
}

@inproceedings{klee,
  author    = {Cristian Cadar and Daniel Dunbar and Dawson Engler},
  title     = {{KLEE}: {U}nassisted and {A}utomatic {G}eneration of {H}igh-{C}overage {T}ests for {C}omplex {S}ystems {P}rograms},
  booktitle = {Proceedings of the 8th USENIX Symposium on Operating Systems Design and Implementation},
  year      = {2008},
  pages     = {209--224},
  publisher = {USENIX Association},
}

@inproceedings{dafny,
  author    = {K. Rustan M. Leino},
  title     = {Dafny: {A}n {A}utomatic {P}rogram {V}erifier for {F}unctional {C}orrectness},
  booktitle = {Proceedings of the 16th International Conference on Logic for Programming, Artificial Intelligence, and Reasoning},
  year      = {2010},
  pages     = {348--370},
  publisher = {Springer},
}

@inproceedings{verus2024sosp,
author = {Lattuada, Andrea and Hance, Travis and Bosamiya, Jay and Brun, Matthias and Cho, Chanhee and LeBlanc, Hayley and Srinivasan, Pranav and Achermann, Reto and Chajed, Tej and Hawblitzel, Chris and Howell, Jon and Lorch, Jacob R. and Padon, Oded and Parno, Bryan},
title = {Verus: A Practical Foundation for Systems Verification},
year = {2024},
isbn = {9798400712517},
publisher = {Association for Computing Machinery},
address = {New York, NY, USA},
url = {https://doi.org/10.1145/3694715.3695952},
doi = {10.1145/3694715.3695952},
booktitle = {Proceedings of the ACM SIGOPS 30th Symposium on Operating Systems Principles},
pages = {438–454},
numpages = {17},
location = {Austin, TX, USA},
series = {SOSP '24}
}

@inproceedings{verus,
  author    = {Andrea Lattuada and Travis Hance and Chanhee Cho and Matthias Brun and Isitha Subasinghe and Yi Zhou and Jon Howell and Bryan Parno and Chris Hawblitzel},
  title     = {Verus: {V}erifying {R}ust {P}rograms using {L}inear {G}host {T}ypes},
  booktitle = {Proceedings of the ACM on Programming Languages (OOPSLA)},
  year      = {2023},
  publisher = {ACM},
}

@inproceedings{lahirie2024evaluating,
  author={Lahirie, Shuvendu K.},
  booktitle={2024 Formal Methods in Computer-Aided Design (FMCAD)}, 
  title={Evaluating LLM-driven User-Intent Formalization for Verification-Aware Languages}, 
  year={2024},
  volume={},
  number={},
  pages={142-147},
  doi={10.34727/2024/isbn.978-3-85448-065-5_19}
}

@article{yang2025autoverus,
author = {Yang, Chenyuan and Li, Xuheng and Misu, Md Rakib Hossain and Yao, Jianan and Cui, Weidong and Gong, Yeyun and Hawblitzel, Chris and Lahiri, Shuvendu and Lorch, Jacob R. and Lu, Shuai and Yang, Fan and Zhou, Ziqiao and Lu, Shan},
title = {AutoVerus: Automated Proof Generation for Rust Code},
year = {2025},
issue_date = {October 2025},
publisher = {Association for Computing Machinery},
address = {New York, NY, USA},
volume = {9},
number = {OOPSLA2},
url = {https://doi.org/10.1145/3763174},
doi = {10.1145/3763174},
journal = {Proc. ACM Program. Lang.},
month = oct,
articleno = {396},
numpages = {29}
}

@inproceedings{chen2024safe,
author = {Chen, Tianyu and Lu, Shuai and Lu, Shan and Gong, Yeyun and Yang, Chenyuan and Li, Xuheng and Misu, Md Rakib Hossain and Yu, Hao and Duan, Nan and Cheng, Peng and Yang, Fan and Lahiri, Shuvendu and Xie, Tao and Zhou, Lidong},
title = {Automated Proof Generation for Rust Code via Self-Evolution},
booktitle = {ICLR 2025},
year = {2024},
month = {October},
url = {https://www.microsoft.com/en-us/research/publication/automated-proof-generation-for-rust-code-via-self-evolution/},
}

@inproceedings{liu2026sharpen,
  title={Sharpen the Spec, Cut the Code: A Case for Generative File System with SYSSPEC},
  author={Liu, Qingyuan and Zou, Mo and Zhang, Hengbin and Du, Dong and Xia, Yubin and Chen, Haibo},
  booktitle={24th USENIX Conference on File and Storage Technologies (FAST 26)},
  pages={291--311},
  year={2026}
}

@misc{opencode,
  author       = {OpenCode Development Team},
  title        = {The open source AI coding agent},
  year         = {2026},
  howpublished = {\url{https://opencode.ai/},},
}

@misc{openrouter,
  author       = {OpenRouter Development Team},
  title        = {OpenRouter},
  year         = {2026},
  howpublished = {\url{https://openrouter.ai/},},
}

@misc{cloc,
  author       = {cloc Development Team},
  title        = {cloc - Count Lines of Code},
  year         = {2026},
  howpublished = {\url{https://github.com/aldanial/cloc},},
}

\end{document}